\documentstyle[12pt]{article}
\def\be{\begin{eqnarray}}
\def\ee{\end{eqnarray}}
\def\ba{\begin{array}}
\def\ea{\end{array}}
\def\p{\phi}

\def\pa{\partial}

\def\M{{\cal M}}
\def\G{{\cal G}}
\def\B{{\cal B}}
\def\A{{\cal A}}
\def\X{{\cal X}}
\def\H{{\cal H}}
\def\S{{\cal S}}

\def\L{{\cal L}}
\def\E{{\cal E}}

\def\T{{\cal T}}

\def\D{^{(D)}}
\def\Z{{\cal Z}}
\def\Q{{\cal Q}}
\def\I{{\cal I}}
\begin{document}
\begin{center}
{\LARGE {Charging Symmetries and Linearizing Potentials\\
\vskip 0.3cm
for Heterotic String in Three Dimensions
}}
\end{center}
\vskip 1.5cm
\begin{center}
{\bf \large {Alfredo Herrera-Aguilar}}
\end{center}
\begin{center}
Joint Institute for Nuclear Research,\\
Dubna, Moscow Region 141980, RUSSIA.\\
e-mail: alfa@cv.jinr.dubna.su
\end{center}
\vskip 0.1cm
\begin{center}
and
\end{center}
\vskip 0.1cm
\begin{center}
{\bf \large {Oleg Kechkin}}
\end{center}
\begin{center}
Institute of Nuclear Physics,\\
Moscow State University, \\
Moscow 119899, RUSSIA, \\
e-mail: kechkin@monet.npi.msu.su
\end{center}
\vskip 1.5cm
\begin{abstract}
Using the Ernst potential formulation we construct all {\it finite}
symmetry
transformations which preserve asymptotics of the bosonic fields of the
$(d+3)$--dimensional low--energy heterotic string theory compactified on a
$d$--torus. We combine all the dynamical variables into a single
$(d+1) \times (d+1+n)$--dimensional matrix potential which
{\it linearly} transforms
under the action of these symmetry transformations in a {\it transparent}
$SO(2,d-1) \times SO(2,d-1+n)$ way, where n is the number of Abelian
vector fields. We formulate the most general solution generation technique
based on the use of these symmetries and show that they form an
{\it invariance group} of the general Israel--Wilson--Perj'es class of
solutions.
\end{abstract}
\newpage
\setcounter{section}{0}
\setcounter{equation}{0}
\renewcommand{\theequation}{1.\arabic{equation}}
\section*{Introduction}
There are some strong arguments that superstring theory provides a correct
quantum description of all fundamental forces including gravity \cite {qg}. 
In the
low--energy limit five $D=10$ consistent perturbative superstring
theories, as well as eleven--dimensional supergravity,
lead to some modifications of General Relativity; these theories are
interrelated by a web of superstring dualities, giving rise to
M--theory \cite {mt}. Usually one considers the bosonic sector
of these {\it superstring gravity models} and extracts from this sector 
supersymmetric solutions, i.e. the on--shell bosonic fields
which do not generate any superfields under the action of supersymmetry
transformations. These supersymmetric, or BPS--saturated solutions,
do not obtain quantum corrections and form a basic tool for the study of the
non--perturbative aspects of superstring theory \cite {bps}.

In this paper we develop new formalism for construction
and symmetry analysis of the bosonic
solutions to the gravity model arising in the framework of
heterotic string (HS) theory. In $D$ dimensions, its effective action reads:
\be
S\D\!=\!\int\!d\D\!x\!\mid\!
G\D\!\mid^{\frac{1}{2}}\!e^{-\p\D}\!(R\D\!+\!
\p\D_{;M}\!\p^{(D);M}
\!-\!\frac{1}{12}\!H\D_{MNP}H^{(D)MNP}\!-\!
\frac{1}{4}F^{(D)I}_{MN}\!F^{(D)IMN}),
\ee
where
\be
F^{(D)I}_{MN}\!=\!\pa_MA^{(D)I}_N\!-\!\pa _NA^{(D)I}_M, \quad
H\D_{MNP}\!=\!\pa_MB\D_{NP}\!-\!\frac{1}{2}A^{(D)I}_M\,F^{(D)I}_{NP}\!+\!
\mbox{{\rm \, cycl \, perms \,\, of} \,\, M,\,N,\,P.}
\nonumber
\ee
Here $G\D_{MN}$ is the metric, $B\D_{MN}$ is the
anti--symmetric Kalb-Ramond field, $\p\D$ is the dilaton and $A^{(D)I}_M$
is the set of $U(1)$ vector fields ($I=1,\,2,\,...,n$). In the consistent
critical case $D=10$ and $n=16$, but we shall leave these parameters
arbitrary in our analysis. Following Maharana, Schwarz \cite {ms} and
Sen \cite {s}, we consider the compactification of this model on a
$D-3=d$--torus. The resulting three--dimensional theory possesses the
$SO(d+1,d+1+n)$ symmetry group \cite {s} (U--duality \cite {u}).

Below we separate U--duality to the gauge and non--gauge sectors. Next,
we fix the gauge (the trivial field asymptotics) and construct a
representation of the theory which linearizes the non--gauge sector.
Transformations of this sector form a {\it charging symmetry} (CS) subgroup.
They generate charged solutions from neutral ones (see \cite {k1} for CS
in the Einstein--Maxwell (EM) theory).

This paper is organized as follows. In Sec. 2 we review the matrix Ernst
potential (MEP) formulation for HS in three dimensions \cite {hk1}. We show
how the EM theory in its ordinary (complex) Ernst potential formulation
\cite {e} arises as some very special case of the HS theory in the
framework of MEP
approach. In Sec. 3 we obtain all the CS transformations in a finite form
using the MEP formulation. After that in Sec. 4 we introduce a pair of 
new matrix variables and show that they transform {\it linearly} under
the action of the all CS transformations. We derive these {\it linearizing
potentials} (LP) for the EM theory using the complex Ernst potential
formulation (the details can be found in the Appendix A), and directly
generalize the result to the general HS theory case.

Next, we combine both LP into a single $(d+1) \times (d+1+n)$ linearizing
potential and show that it transforms in a manifest
$SO(2, d-1) \times SO(2, d-1+n)$ form. This allow us to establish the CS
group structure (the non--trivial structure of its $so(2, d-1+n)$ subalgebra
is studied in the Appendix B).
After that we construct from this single LP one charging
symmetry invariant (CSI) which is closely related to the charge quadratic
function vanishing for the BPS--saturated fields.

Sec. 5 contains an investigation of the HS fields with a linear dependence
between the linearizing potentials. First, we show that such fields are
{\it invariant} under the action of charging symmetries, so that this
class of fields
cannot be generalized using the CS transformations. By setting the CSI
to zero we get a restriction on this class. The corresponding subclass
coincides with the general class of Israel--Wilson--Perj'es (IWP)
solutions of
the three--dimensional heterotic string theory. We show that the
restriction mentioned above is invariant under the CS transformations;
thus, the IWP solutions form a CS invariant class of solutions.

In Sec. 6 we formulate the most general technique for generation of new
solutions based on the use of charging symmetries. We show, how starting
from the pure Kaluza--Klein theory one can generate all the massless
fields of the
bosonic string theory and, next, the full heterotic string theory sector.

We conclude this work with a discussion about the charging symmetry subgroup
and charging symmetry invariant in the non--perturbative regime of the
heterotic string theory.
\setcounter{section}{0}
\setcounter{equation}{0}
\renewcommand{\theequation}{2.\arabic{equation}}
\section*{Matrix Ernst Potentials}
Matrix Ernst potentials contain all information about the dynamical variables
for the heterotic string theory reduced to three dimensions. These variables
consist of (\cite{ms}-\cite{s})

a) scalar fields
\be
G\!=\!\left (G_{pq}\!=
\!G\D_{p+3,q+3}\right ),\quad
B\!=\!\left ( B_{pq}\!=
\!B\D_{p+3,q+3}\right ) ,\quad
A\!=\!\left ( A^I_p\!=
\!A^{(D)I}_{p+3}\right ) ,\quad
\p\!=\!\p\D\!-\!\frac{1}{2}{\rm ln \,| det}\,G|,
\nonumber
\\
\ee
where the subscripts $p,q=1,2,...,d$.

b)tensor fields
\be
g_{\mu\nu}\!=\!e^{-2\p}\left(G\D_{\mu\nu}\!-\!G\D_{p+3,\mu}G\D_{q+3,\nu}G^{pq}
\right),\,\,\,
B_{\mu\nu}\!=\!B\D_{\mu\nu}\!-\!4B_{pq}A^p_{\mu}A^q_{\nu}\!-\!
2\left(A^p_{\mu}A^{p+d}_{\nu}-A^p_{\nu}A^{p+d}_{\mu}\right),
\nonumber
\\
\ee
(we put $B_{\mu\nu}=0$ to remove the effective three--dimensional
cosmological constant from our consideration \cite {sugra}).

c)vector fields $A^{(a)}_{\mu}=
\left((A_1)^p_{\mu},(A_2)^{p+d}_{\mu},(A_3)^{2d+I}_{\mu}\right)$
($a=1,...,2d+n$)
\be
(A_1)^p_{\mu}\!=\!\frac{1}{2}G^{pq}G\D_{q+3,\mu},\quad
(A_3)^{I+2d}_{\mu}\!=\!-\frac{1}{2}A^{(D)I}_{\mu}\!+\!A^I_qA^q_{\mu},\quad
(A_2)^{p+d}_{\mu}\!=\!\frac{1}{2}B\D_{p+3,\mu}\!-\!B_{pq}A^q_{\mu}\!+\!
\frac{1}{2}A^I_{p}A^{I+2d}_{\mu}.
\nonumber
\\
\ee
These variables form a complete set of the $N=8$ supergravity in the
critical case \cite {sugra}.

In three dimensions all vector fields can be dualized on--shell:
\begin{eqnarray}
\nabla\times\overrightarrow{A_1}&=&\frac{1}{2}e^{2\p}G^{-1}
\left(\nabla u+(B+\frac{1}{2}AA^T)\nabla v+A\nabla s\right),
\nonumber                          \\
\nabla\times\overrightarrow{A_3}&=&\frac{1}{2}e^{2\p}
(\nabla s+A^T\nabla v)+A^T\nabla\times\overrightarrow{A_1},
\nonumber                            \\
\nabla\times\overrightarrow{A_2}&=&\frac{1}{2}e^{2\p}G\nabla v-
(B+\frac{1}{2}AA^T)\nabla\times\overrightarrow{A_1}+
A\nabla\times\overrightarrow{A_3}.
\end{eqnarray}
The resulting three--dimensional theory describes the scalars
$G$, $B$, $A$ and $\p$ and
pseudoscalars $u$, $v$ and $s$ coupled to the metric $g_{\mu\nu}$.

We define the {\it matrix} \, Ernst potentials as follows:
\be
\X=
\left(
\ba{cc}
-e^{-2\p}+v^TXv+v^TAs+\frac{1}{2}s^Ts&v^TX-u^T \cr
Xv+u+As&X
\ea
\right), \quad {\rm and} \quad
\A=\left(
\ba{c}
s^T+v^TA \cr
A
\ea
\right),
\ee
where $X=G+B+\frac{1}{2}AA^T$. They are of the dimensions
$(d+1) \times (d+1)$ and $(d+1) \times n$
correspondingly. In terms of MEP the effective three--dimensional theory
takes the form:
\be
S^{(3)}\!&&=
\!\int\!d^3x\!\mid g\mid^{\frac{1}{2}}\!\{\!-\!R\!+\!{\rm Tr}\,[
\frac{1}{4}\left(\nabla \X\!-\!\nabla \A\A^T\right)\!\G^{-1}
\!\left(\nabla \X^T\!-\!\A\nabla \A^T\right)\!\G^{-1}
\!+\!\frac{1}{2}\nabla \A^T\G^{-1}\nabla \A]\}
\nonumber\\
&&=\int d^3x \mid g\mid^{\frac{1}{2}}\{-R+\L_{{}_{HS}}\},
\ee
where
\be
\G=\frac{1}{2}\left(\X+\X^T-\A\A^T\right).
\ee

It is well--known that the four--dimensional Einstein--Maxwell theory,
being reduced to three
dimensions, allows a similar formulation using two {\it complex} Ernst
potentials
$E$ and $F$ \cite {e}. Let us rewrite them in the less conventional form
\be
\X_{{}_{EM}}={\rm Re}E+\sigma_2\,{\rm Im}E, \quad
\A_{{}_{EM}}={\rm Re}F+\sigma_2\,{\rm Im}F, \quad
{\rm where} \quad
\sigma_2=
\left(
\ba{cc}
0&-1 \cr
1&0
\ea
\right).
\ee
We can treat these matrices as the matrix Ernst potentials (2.4) of the $D=4$
theory (1.1) with $\phi^{(4)}=B^{(4)}_{MN}=0$. Then we conclude that $n=2$
and
\be
s^1=A^2={\rm Re}F, \quad -s^2=A^1={\rm Im}F.
\ee
Note, that $s^I$ $(I=1,2)$ describe the magnetic potentials, whereas
$A^I$ are the electric ones. Thus, two Maxwell fields arising in the
framework of the representation (2.5)--(2.8) occur to be mutually conjugated
(i.e. $F_{MN}^{(4)2}=\tilde F_{MN}^{(4)1}$ in four dimensions).
Next, for the single extra metric component one has:
\be
G=
\frac {1}{2}\left (E+\bar E-F\bar F\right ) \equiv f,
\ee
whereas the rotational potential $u$ becomes equal to ${\rm Im}E$. Taking
into account that $\G=G$, and substitute Eqs. (2.8), (2.10) into Eq. (2.6),
we obtain
\be
\L_{{}_{EM}}=\frac{1}{2f^2}\left |\nabla E-\bar F\nabla F\right |^2+
f^{-1}
\left |\nabla F\right |^2.
\ee
This is the matter Lagrangian of the three--dimensional EM theory \cite {e}.
Thus, our MEP
formulation (2.5)--(2.7) of the HS theory includes the EM theory as a
special case.

One can see that formally the transition from the HS theory (2.6) to the EM
theory (2.11) is defined by the substitution
\be
\X\rightarrow E, \quad \A\rightarrow F
\ee
together with the replacement 
\be
{\it matrix\,\, transposition}\quad\rightarrow\quad 
{\it complex\,\, conjugation}.
\ee
The inverse map (EM $\rightarrow$ HS) allow us to generalize the results
for the EM theory, obtained using its complex representation, to the ones for
the HS theory in the MEP formulation. This becomes possible if the
matrix--valued generalized relation can be written in a form which is free
of the matrix multipliers ordering problem. 
\setcounter{section}{0}
\setcounter{equation}{0}
\renewcommand{\theequation}{3.\arabic{equation}}
\section*{Charging Symmetries}
The complete symmetry group (U--duality \cite {u}) of the
effective three--dimensional heterotic
string theory (2.6) is $SO(d+1,d+1+n)$ \cite {s}.
Its action on the matrix Ernst potentials had been established in the our
previous
work \cite {hk1}. There was shown that one discrete transformation,
which is closely related to the so--called strong--weak coupling
duality (we will call this transformation `SWCD' \cite {swcd})
plays a crucial role in the `web' of HS
symmetries (or dualities). Actually, the SWCD transformation
\be
\X\rightarrow \X^{-1} \qquad \A\rightarrow -\X^{-1}\A.
\ee
maps the $\X$--shift symmetry
\be
\X\rightarrow\X+\lambda_{\X}, \qquad \A\rightarrow\A
\quad {\rm where} \quad \lambda_{\X}^T=-\lambda_{\X}
\ee
into the Ehlers transformation
\be
\X\rightarrow\left(1+\X\lambda_{\E}
\right)^{-1}\X, \qquad
\A\rightarrow\left(1+\X\lambda_{\E}
\right)^{-1}\A, \quad {\rm where} \quad \lambda_{\E}^T=-\lambda_{\E};
\ee
whereas the $\A$--shift symmetry
\be
\A\rightarrow\A+\lambda_{\A}, \quad \X\rightarrow\X+
\A\lambda_{\A}^T+\frac{1}{2}\lambda_{\A}\lambda_{\A}^T
\ee
maps into the Harrison transformation
\be
\A\rightarrow\left(1-\A\lambda_{\H}^T+\frac{1}{2}\X
\lambda_{\H}\lambda_{\H}^T\right)^{-1}\left(\A-\X\lambda_{\H}\right),
\quad
\X\rightarrow\left(1-\A\lambda_{\H}^T+\frac{1}{2}
\X\lambda_{\H}\lambda_{\H}^T\right)^{-1}\X.
\ee
The remaining symmetries, i.e. the electric--magnetic rotation
\be
\A\rightarrow\A\T, \qquad   \X\rightarrow\X, \quad {\rm where} \quad \T\T^T=1
\ee
and the scaling transformation
\be
\X\rightarrow\S^T\X\S, \qquad \A\rightarrow\S^T\A
\ee
are the SWCD invariants (with $\T\rightarrow \T$ and
$\S\rightarrow (\S^T)^{-1}$). Thus, U--duality of the three--dimensional
effective heterotic
string theory consists of two doublets and two singlets of the strong--weak
coupling duality transformation.

Now let us consider the arbitrary constant potentials
$\X=\X_{{}_{\infty}}$ and $\A=\A_{{}_{\infty}}$,
which can have interpretation of the asymptotics near the spatial infinity.
Applying the $\A$--shift symmetry with $\lambda_{\A}=-\A_{{}_{\infty}}$ we annihilate
the $\A$--potential. Next, using the $\X$--shift with $\lambda_{\X}=
(\X_{{}_{\infty}}^T-\X_{{}_{\infty}})/2$ we remove the antisymmetric part of $\X$. Finally, the
scaling with
\be
\S=
\left(
\ba{cc}
e^{-\p_{{}_{\infty}}}&0\cr
-e^{-\p_{{}_{\infty}}}v^T_{{}_{\infty}}&\gamma^{-1}_{{}_{\infty}}
\ea
\right),
\ee
where $G_{{}_{\infty}}=\gamma^T_{{}_{\infty}}\sigma\gamma_{{}_{\infty}}$,
leads $\X$ to the its trivial form
\be
\Sigma=
\left(
\ba{cc}
-1&0\cr
0&\sigma
\ea
\right).
\ee
Here $\sigma=diag(-1,1,...,1)$ is the tangent metric for $G_{{}_{\infty}}$,
whereas
$\gamma_{{}_{\infty}}$ is the tetrad matrix. Thus, U--duality contains gauge
transformations which can be used for the removing of the all field
asymptotics.
Conversely, one can apply these `dressing' transformations to obtain arbitrary
asymptotics for the originally asymptoticaly--free field
configuration. In the rest part of the paper we fix the gauge and put
$\X_{{}_{\infty}}=\Sigma$ and $\A_{{}_{\infty}}=0$.

The transformations preserving fixed asymptotics form a subgroup, which
we call `charging' by the reasons given below. The electric--magnetic
rotation belongs to this subgroup; another
representative is related to the scaling transformation. Actually, the
trivial $\A$--asymptotic preserves for the arbitrary value of $\S$; however
only the scalings constrained by
\be
\S^T\Sigma\S=\Sigma
\ee
do not change the chosen $\X$--asymptotic. Thus, the group of charging
symmetries contains the $SO(2,d-1)$ subgroup of the scaling symmetry.

One can see that the Ehlers transformation with the arbitrary non--trivial
parameter $\lambda_{\E}$ moves the asymptotical value
$\X_{{}_{\infty}}=\Sigma$.
However, some combination of
the Ehlers transformation with the special $\X$--shift and scaling
duality belongs to the charging symmetry subgroup. Actually, let us
suppose that the Ehlers transformation with the arbitrary antisymmetric
parameter $\lambda_{\X}$ is applied to the matrices
$\X_{{}_{\infty}}=\Sigma$ and $\A_{{}_{\infty}}=0$. Then the
value of $\A_{{}_{\infty}}$ remains trivial,
whereas $\X_{{}_{\infty}}$ becomes changed. Then, to remove the
antisymmetric part
of new $\X_{{}_{\infty}}$, we perform the $\X$--shift transformation with
$\lambda_{\X}=(1+\Sigma\lambda_{\E})^{-1}\Sigma\lambda_{\E}\Sigma
(1-\lambda_{\E}\Sigma)^{-1}$. Finally, we transform the resulting
$\X_{{}_{\infty}}$--asymptotic to $\Sigma$ using the scaling (3.7) with
$\S=1-\lambda_{\E}\Sigma$. The resulting transformation has the form
\be
&&\X\rightarrow\left(1+\Sigma\lambda_{\E}\right)
\left(1+\X\lambda_{\E}\right)^{-1}\X(1-\lambda_{\E}\Sigma)+
\Sigma\lambda_{\E}\Sigma,
\nonumber
\\
&&\A\rightarrow
\left(1+\Sigma\lambda_{\E}\right)
\left(1+\X\lambda_{\E}\right)^{-1}\A.
\ee
We call this symmetry {\it normalized Ehlers transformation} (NET).

The similar `normalization' procedure can be performed for the Harrison
transformation with the arbitrary matrix parameter $\lambda_{\H}$. First, we
remove the $\A$--asymptotic generated by the Harrison transformation
using the $\A$--shift symmetry with $\lambda_{\A}=\left(1+\Sigma\frac{1}{2}
\lambda_{\H}\lambda_{h}^T\right)^{-1}\Sigma\lambda_{\H}$. Second, we
transform the `broken' $\X$--asymptotic to $\Sigma$ using the scaling with
$\S=1+\frac{1}{2}\Sigma\lambda_{\H}\lambda_{\H}^T$. The resulting {\it
normalized Harrison transformation} (NHT) is defined by the formulae
\be
&&\A\rightarrow\left(1+\frac{1}{2}\Sigma\lambda_{\H}\lambda_{\H}^T\right)
\left(1-\A\lambda_{\H}^T+\frac{1}{2}\X\lambda_{\H}\lambda_{\H}^T\right)^{-1}
\left(A-\X\lambda_{\H}\right)+\Sigma\lambda_{\H},
\\
&&\X\rightarrow\left(1+\frac{1}{2}\Sigma\lambda_{\H}\lambda_{\H}^T\right)
\left(1-\A\lambda_{\H}^T+\frac{1}{2}\X\lambda_{\H}\lambda_{\H}^T\right)^{-1}
\left[\X+\left(\A-\frac{1}{2}\X\lambda_{\H}\right)\lambda_{\H}^T\Sigma\right]
+\frac{1}{2}\Sigma\lambda_{\H}\lambda_{\H}^T\Sigma.
\nonumber
\ee

The number of the dressing symmetries is equal to the number of
the dynamical variables of HS theory
(2.6) , i.e. to $(d+1)(d+1+n)$. The electric--magnetic
rotations
(3.6) form the SO(n) subgroup defined by $n(n-1)/2$ parameters,
whereas the $SO(2,d-1)$ scaling subgroup (3.7), as well as NET (3.11), gives
$(d+1)d/2$ parameters.
Finally, the NHT is related to the parameter matrix with
$(d+1)n$ elements. Thus, all the established 
transformations from the CS subgroup, being independent, are constructed
from $(d+1)(d+n)+n(n-1)/2$ parameters. Then the
common
number of independent dressing and charging transformations becomes equal to 
$\left[2(d+1)+n\right]\left[2(d+1)+n-1\right]/2$, i.e. to the number
of parameters of the whole U--duality group $SO(d+1,d+1+n)$. From this it
follows that we have found all the gauge (dressing) transformations as
well as all the non--gauge (charging) symmetries. Thus, the CS subgroup
consists of the electric--magnetic rotation (3.6), the scaling (3.7)
restricted by
Eq. (3.10), and the normalized Ehlers (3.11) and Harrison (3.12)
transformations.

In the EM theory case the $\X$--shift, Ehlers, electric--magnetic and
scaling transformations become one--parametric symmetries;
they are defined by the linear combinations of the two--dimensional
unit matrix and  $\sigma_2$ (see Eqs. (3.2), (3.3), (3.6), (3.7) and (3.10)).
To preserve
the same structure of the matrix Ernst potentials (2.8) under the
action of the
remaining two symmetries, we must suppose that
the $\A$--shift and Harrison transformations are defined by 
two--parametric
linear combinations of the unit and $\sigma_2$ matrices. 
Finally, the general symmetry group (`U--duality') of the
EM theory in three dimensions occurs to be eight--parametric. This really
takes place because this group is isomorphic to $SU(1,2)$ \cite {k2}.

Using the substitution (2.12) it is easy to rewrite all the symmetries in the
conventional form of the complex Ernst potentials (the matrix $\sigma_2$
must be changed
to the imaginary unit $i$). Below we will need only in the CS
transformations;
they can be obtained from Eqs. (3.6), (3.7), (3.10), (3.11) and (3.12)
using the additional
replacement
\be
\Sigma\rightarrow -1
\ee
(we consider the time compactification for definiteness )
\be
E\rightarrow E, \quad F\rightarrow e^{i\alpha}F;
\quad \left ( {\rm EMT}\right )
\ee
\be
E\rightarrow \frac {E+i\epsilon}{1+i\epsilon E},
\quad
F\rightarrow \frac {1-i\epsilon}{1+i\epsilon E}F;
\quad \left ( {\rm NET}\right )
\ee
\be
E\rightarrow \frac {E+\frac {1}{2}\left |\lambda_{\H}\right |^2
-\bar \lambda_{\H}F}
{1-\bar \lambda_{\H}F+\frac {1}{2}
\left |\lambda_{\H}\right |^2E},
\quad
F\rightarrow \frac
{\left (1+\frac {1}{2}\left |\lambda_{\H}\right |^2\right )F-
\lambda_{\H}\left ( E+1\right )}
{1-\bar \lambda_{\H}F+\frac {1}{2}
\left |\lambda_{\H}\right |^2E},
\quad \left ( {\rm NHT}\right )
\ee
In Eq. (3.16) the parameter $\lambda _{\H}$ is complex; the other two
parameters $\alpha$ and $\epsilon$ are real. From Eqs. (3.6), (3.7) and
(3.10) it follows
that NST and EMT coincide in the case of a single Maxwell field. We call
the corresponding transformation `electric--magnetic transformation'
because this name remains natural for the case of $n>1$.
\setcounter{section}{0}
\setcounter{equation}{0}
\renewcommand{\theequation}{4.\arabic{equation}}
\section*{Linearizing Potentials}
One can see that the electric--magnetic rotation and scaling act as
linear
transformations on the matrix Ernst potentials $\X$ and $\A$, whereas
the normalized Ehlers and Harrison transformations are some
fractional--linear
maps. In this section we introduce new matrix potentials $\Z_1$ and $\Z_2$
which linearly transform under the action of the {\it all} CS
transformations, i.e. they form a possible set of CS linearizing
potentials.

In \cite {hk2} we have found the linearizing potentials for the special case
of $d=n=1$ (the so--called Einstein--Maxwell dilaton--axion
(EMDA) theory). This theory allows a K\"ahler representation using 
three complex Ernst potentials \cite {gk}. First, we constructed a linear
realization of
the CS subgroup on the set of three over complex variables. Second, we
computed all the commutation relations between the CS generators in the both
representations and, finally, we identify the generators accordingly to
their
commutation relations. This identification gave us differential
equations of the first order which define the unknown functional relations
between the Ernst and linearizing potentials. The result of \cite {hk2} can
easily
be extended to the case of $d=1$ and arbitrary $n$. However, the following
extension to the case of $d>1$ seems impossible because the
corresponding K\"ahler formulation for the theory is absent.

To construct the linearizing potentials for the arbitrary $d$ and $n$
(including the critical case of $d=7$,\, $n=16$), first we will find them
for the EM theory in its complex representation. After that, using the map
(2.12), (3.13) in its inverse form, we will obtain LP in the general case.

In the Appendix A one can find the details of the LP derivation for the 
Einstein--Maxwell theory using the strategy formulated above for the EMDA
theory. The result is:
\be
\Z_1=2\left(E-1\right)^{-1}+1, \quad \Z_2=\sqrt 2\left(E-1\right)^{-1}F
\qquad \left( {\rm EM \quad theory}\right) 
\ee
One can see that these equations admit the straightforward HS
generalization, because the problem of the matrix multiplier ordering
does not arise. Then
\be
\Z_1=2\left(\X+\Sigma\right)^{-1}-\Sigma, \quad \Z_2=\sqrt 2\left(\X+\Sigma
\right)^{-1}\A
\qquad \left( {\rm HS \quad theory}\right). 
\ee
One can prove that the inverse relations can be obtained using the
replacements $\Z_1\leftrightarrow\X$ and $\Z_2\leftrightarrow\A$. This
means that the inverse functions of the functions (4.2) coincide
with them. Actually, this remarkable property fixes the multiplier
`$\sqrt 2$'.

We have supposed that the fractional--linear functions of Eq. (4.2) define
the linearizing potentials 
for the HS theory. To verify this, it is necessary to rewrite all the CS
transformations using the $(\Z_1,\Z_2)$--representation. The result is:
\be
&&\Z_1\rightarrow\Z_1, \quad \Z_2\rightarrow\Z_2\T;
\qquad \left( {\rm EMT}\right)
\ee
\be
&&\Z_1\rightarrow\S^{-1}\Z_1(\S^T)^{-1}, \quad \Z_2\rightarrow\S^{-1}\Z_2;
\qquad \left( {\rm scaling
}\right)
\ee
\be
&&\Z_1\rightarrow\Z_1
\left(1-\Sigma\lambda_{\E}\right)
\left(1+\Sigma\lambda_{\E}\right)^{-1},
\quad \Z_2\rightarrow\Z_2;
\qquad \left( {\rm NET
}\right)
\ee
\be
&&\Z_1\rightarrow\Z_1
\left(1-\frac{1}{2}\Sigma\lambda_{\H}\lambda_{\H}^T\right)
\left(1+\frac{1}{2}\Sigma\lambda_{\H}\lambda_{\H}^T\right)^{-1}
-\sqrt 2\Z_2\lambda_{\H}^T
\left(1+\frac{1}{2}\Sigma\lambda_{\H}\lambda_{\H}^T\right)^{-1},
\\
&&\Z_2\rightarrow\Z_2\left[
1-\lambda_{\H}^T
\left(\Sigma+\frac{1}{2}\lambda_{\H}\lambda_{\H}^T\right)^{-1}\lambda_{\H}
\right]
+\sqrt 2\Z_1 
\left(\Sigma+\frac{1}{2}\lambda_{\H}\lambda_{\H}^T\right)^{-1}\lambda_{\H},
\qquad \left( {\rm NHT
}\right)
\nonumber
\ee
where `EMT' denotes the electric--magnetic transformation. From these
formulae we see that the linearization really takes place.
Thus, $\Z_1$ and $\Z_2$ actually form a set of linearizing potentials of
the charging symmetry subgroup of the three--dimensional heterotic string
theory (2.6).

The linearizing potentials allow us to clarify the CS subgroup structure.
Actually, the defining NET matrix
\be
\E=\left(1-\Sigma\lambda_{\E}\right)
\left(1+\Sigma\lambda_{\E}\right)^{-1}
\ee
satisfies the
$SO(2,d-1)$ group relation $\E^T\Sigma\E=\Sigma$. Conversely,
it is easy to see that any $SO(2,d-1)$--matrix can be written in the form
of Eq. (4.7). Thus, the
normalized Ehlers transformation forms the $SO(2,d-1)$ CS subgroup. Next,
NET can be explored for some `normalization' of the scaling symmetry
subgroup. Actually,
these both transformations are represented by the matrices of the
$SO(2,d-1)$ group.
We define the {\it normalized scaling transformation} (NST) as result
of the action of the scaling with $\S\in SO(2,d-1)$ and NET with
$\E=\S^T$:
\be
&&\Z_1\rightarrow\S^{-1}\Z_1, \quad \Z_2\rightarrow\S^{-1}\Z_2. 
\qquad \left( {\rm NST
}\right)
\ee
The significance of the scaling subgroup normalization is related with the
fact that NST commutes with
all other CS transformations, because NST acts as the {\it left multiplier}
on $\Z_1$ and $\Z_2$ whereas EMT, NET
and NHT are the {\it right multipliers} (see Eqs. (4.3), (4.5), (4.6) and
(4.8)).

To analyse the group structure of the right multiplier
sector (the single left multiplier forms the $SO(2,d-1)$ subgroup)
we will need in one general CS invariant.
This invariant can be `extracted' from the Lagrangian $\L_{{}_{HS}}$
(see Eq. (2.6)). Actually, let us consider the asymptotically trivial fields
with the non--zero Coulomb terms: 
\be
\Z_1=\frac{\Q_1}{r}+o\left(\frac{1}{r}\right), \quad
\Z_2=\frac{\Q_2}{r}+o\left(\frac{1}{r}\right),
\ee
where $\Q_1$ and $\Q_2$ are the charge matrices and $r$ tends to the spatial
infinity. Then, using the inverse to Eqs. (4.2) relations we obtain that
\be
\L_{{}_{HS}}=\frac{1}{r^4}
{\rm Tr}\left \{ \Q_1^T\Sigma\Q_1\Sigma+\Q_2^T\Sigma\Q_2\right \}
+o\left(\frac{1}{r^4}\right).
\ee
The quadratic charge combination $\I=
{\rm Tr}\left \{ \Q_1^T\Sigma\Q_1\Sigma+\Q_2^T\Sigma\Q_2\right \}$ is the 
CS invariant. It is so because $\L_{{}_{HS}}$ is the CS invariant and all
its
terms related to the $1/r$ power expansion are also CS invariants.
Now it is
useful to introduce the $(d+1)\times (d+1+n)$ charge matrix $\Q=(\Q_1,\Q_2)$
and to rewrite $\I$ in the form
\be
\I(\Q)={\rm Tr}\left \{\Q^T\Sigma\Q\,\Xi\right \}, \quad {\rm where} \quad
\Xi=
\left(
\ba{cc}
\Sigma&0\cr
0&1
\ea
\right).
\ee
Let us now introduce the $(d+1)\times (d+1+n)$ linearizing potential
$\Z=(\Z_1,\Z_2)$. Then, from Eq. (4.9) it follows that the charge and
linearizing potential matrices have the same transformation properties. Thus,
the function
\be
\I(\Z)={\rm Tr}\left \{\Z^T\Sigma\Z\,\Xi\right \}
\ee
must be a charging symmetry invariant.

All the charging symmetry transformations can be rewritten in the form
\be
\Z\rightarrow\G^{{}^{left}}_i\Z\G^{{}^{right}}_i,
\ee
where $i=$ NST, EMT, NET or NHT
and an explicit
form of the matrices $\G^{{}^{left}}_i$ and $\G^{{}^{right}}_i$
can be obtained from Eqs. (4.3), (4.5), (4.6) and (4.8). The only
non--unit matrices are:
\be
&&\G^{{}^{left}}_{{}_{\rm NST}}=
\left(
\ba{cc}
\S^{-1}&0\cr
0&1
\ea
\right),
\qquad
\G^{{}^{right}}_{{}_{\rm EMT}}=
\left(
\ba{cc}
1&0\cr
0&\T
\ea
\right),
\qquad
\G^{{}^{right}}_{{}_{\rm NET}}=
\left(
\ba{cc}
\E&0\cr
0&1
\ea
\right),
\nonumber
\\
\, \nonumber\\
&&\G^{{}^{right}}_{{}_{\rm NHT}}=
\left(
\ba{cc}
\left(1-\frac{1}{2}\Sigma\lambda_{\H}\lambda_{\H}^T\right)
\left(1+\frac{1}{2}\Sigma\lambda_{\H}\lambda_{\H}^T\right)^{-1}
&
\sqrt 2
\left(\Sigma+\frac{1}{2}\lambda_{\H}\lambda_{\H}^T\right)^{-1}\lambda_{\H}
\cr
-\sqrt 2\lambda_{\H}^T
\left(1+\frac{1}{2}\Sigma\lambda_{\H}\lambda_{\H}^T\right)^{-1}
&
1-\lambda_{\H}^T
\left(\Sigma+\frac{1}{2}\lambda_{\H}\lambda_{\H}^T\right)^{-1}\lambda_{\H}
\ea
\right).
\ee
To preserve $\I(\Z)$ all the `left' transformations must satisfy the
relation
\be
(\G^{{}^{left}}_i)^T\,\Sigma\,\G^{{}^{left}}_i\,=\,\Sigma,
\ee
whereas all the `right' transformations must be constrained accordingly to
\be
\G^{{}^{right}}_i\,\Xi\,\left(\G^{{}^{right}}_i\right)^T\,=\,\Xi.
\ee
The first relation really takes place (see Eq. (3.10)); this means that
$\G^{{}^{left}}_i\in SO(2,d-1)$. The direct
substitution of the `right' matrices from Eqs. (4.14) to Eq. (4.16) gives
the
identity. Taking into account the explicit form of $\Xi$ we conclude that
all the transformations $\G^{{}^{right}}_i\in SO(2,d-1+n)$.

Now let us note that the common number of independent parameters of 
EMT, NET and NHT is $(d+1+n)(d+n)/2$, i.e. the same one
as for the $SO(2, d-1+n)$ group. Moreover, if we consider the infinitesimal
transformations $\G^{{}^{right}}_i=1+\Gamma^{{}^{right}}_i$ and
compute $\Gamma^{{}^{right}}
=\sum_i \Gamma_i$, we obtain
\be
\Gamma^{{}^{right}}=
\left(
\ba{cc}
-2\Sigma\lambda_{\E}&\sqrt 2\Sigma\lambda_{\H}\cr
-\sqrt 2\lambda_{\H}^T&\tau
\ea
\right),
\ee
where $\tau=\T-1$. This matrix is the {\it general} solution of the equation
$\left (\Gamma^{{}^{right}}\right )^T=\,-\,\Xi\,\Gamma^{{}^{right}}\,\Xi$,
which defines the $so(2,d-1+n)$ algebra (its structure is discussed
in the Appendix B).
From this we conclude that the general `right' transformation
matrix is the general matrix of the group
$SO(2,d-1+n)$. It can be constructed as the product of the all
$\G^{{}^{right}}_i$ matrices multiplied in an arbitrary order.

Thus, we have established the following {\it simplest} form of the charging
symmetry transformations for the effective three--dimensional
heterotic string theory at low energies:
\be
\Z\rightarrow\G^{{}^{left}}\Z\,\G^{{}^{right}},
\quad {\rm where} \quad \G^{{}^{left}}\in SO(2,d-1) \quad {\rm and}
\quad \G^{{}^{right}}\in SO(2,d-1+n).
\ee
The charging symmetry subgroup occurs to be $\G^{{}^{left}}\times
\G^{{}^{right}}$, i.e. it is isomorphic to
$SO(2,d-1)\times SO(2,d-1+n)$.
It is important to note that the transformation of the charge matrix
$\Q$ can be obtained from Eq. (4.18) using the replacement
$\Z\rightarrow\Q$.

The strong--weak
coupling duality transformation in terms of the linearizing potential
$\Z$ takes the form:
\be
\Z\rightarrow -\Sigma\,\Z\,\Xi.
\ee
From Eqs. (4.18) and (4.19) it follows that the SWCD acts on
the subgroups $\G^{{}^{left}}$ and $\G^{{}^{right}}$ in the following way:
\be
&&\G^{{}^{left}}\rightarrow\Sigma\G^{{}^{left}}\Sigma
\nonumber\\
&&\G^{{}^{right}}\rightarrow\Xi\G^{{}^{right}}\Xi
\ee
One can see that these maps preserve the group relations (4.15) and (4.16).
Thus, both $\G^{{}^{left}}$ and $\G^{{}^{right}}$ are the SWCD--invariant.
This means, that the whole CS subgroup is also SWCD invariant.
Taking into
account that all the dressing symmetries do not possess this property we
obtain the following alternative definition of the CS subgroup:
{\it the charging
symmetry subgroup is the maximal subgroup of the U--duality which is
invariant
under the action of the SWCD transformation}.
\setcounter{section}{0}
\setcounter{equation}{0}
\renewcommand{\theequation}{5.\arabic{equation}}
\section*{Invariant Fields}
Charging symmetries act as {\it linear homogeneous transformations} 
on the linearizing potentials. Let us consider the simplest 
{\it linear homogeneous field configuration}
\be
\Z_2=\Z_1\B,
\ee
where $\B$ is the constant $(d+1)\times n$ matrix. It is easy to see that
this configuration
preserves its form under the action of charging symmetries. Thus,
Eq. (5.1) describes {\it the charging symmetry invariant class of fields}.
The transformation lows for the matrix $\B$ are:
\be
\B\rightarrow\B \quad {\rm for \,\, NST}, \qquad
\B\rightarrow\B\T \quad {\rm for \,\, EMT}, \qquad
\B\rightarrow\E^{-1}\B  \quad {\rm for \,\, NET};
\ee
whereas for NHT one has
\be
\B\rightarrow\left (1+
\frac{1}{2}\Sigma \lambda_{\H}\lambda_{\H}^T\right )
\left (1-\frac{1}{2}\Sigma \lambda_{\H}\lambda_{\H}^T-
\right.
&\sqrt 2&
\left.
\phantom {\frac{1}{2}}\!\!\!
\B\lambda_{\H}^T\right )^{-1}
\left \{
\B\left [1-\lambda_{\H}^T\left (\Sigma +\frac{1}{2}\lambda_{\H}\lambda_{\H}^T
\right )^{-1}\lambda_{\H}\right ]
\right.
\nonumber\\
+&\sqrt 2&
\left.
\left (\Sigma +\frac{1}{2}\lambda_{\H}\lambda_{\H}^T
\right )^{-1}\lambda_{\H}
\right \}
\ee
Eqs. (5.2) and (5.3) describe the realization of the CS subgroup on the
matrix $\B$
(which is trivial for $\G^{{left}}$ and non--trivial for $\G^{{right}}$).

For the EM theory in its complex representation $\B$ becomes the complex
parameter. EMT and NET occur to be equal; they are given by the factor
$\T=\E^{-1}=\exp {(i\delta)}$. The corresponding generator is
\be
\Gamma_{{}_{NET}}=i\B\partial_{\B}.
\ee
Next, NHT consists of two
real transformations; decomposing its complex parameter to the real and
imaginary parts,
$\lambda_{\H}=
\frac {1}{2\sqrt 2}\left (\lambda_{{\H}_1}+i\lambda_{{\H}_2}\right )$,
we obtain:
\be
&&\Gamma_{{}_{{NHT}_1}}=\frac {1}{2}\left (\B^2-1\right )\partial_{\B},
\nonumber
\\
&&\Gamma_{{}_{{NHT}_2}}=-\frac {i}{2}\left (\B^2+1\right )\partial_{\B}
\ee
(we write down only the holomorphic terms).
Now it is easy to verify that the generators (5.4)--(5.5) satisfy the
commutation relations of the $su(1,1)$ algebra:
\be
\left [\Gamma_{{}_{{NHT}_1}},\,\Gamma_{{}_{NET}} \right ]=
\Gamma_{{}_{{NHT}_2}},
\quad
\left [\Gamma_{{}_{NET}},\,\,\Gamma_{{}_{{NHT}_2}} \right ]=
\Gamma_{{}_{{NHT}_1}},
\quad
\left [\Gamma_{{}_{{NHT}_1}},\,\,\Gamma_{{}_{{NET}_2}} \right ]=
-\Gamma_{{}_{NET}}.
\ee

Next, in the general HS theory case the charging symmetry invariant,
being computed for the fields (5.1), is equal
to
$\I(\Z)={\rm Tr}\left [\Z_1^T\Sigma\Z_1\left (\B\B^T+\Sigma\right )\right ]$;
it vanishes if
\be
\B\B^T=-\Sigma.
\ee
It is easy to see that this restriction is invariant itself under the
action of NST, EMT and NET. After some algebraic manipulations one can prove
that NHT (5.3) also preserves Eq. (5.7). Thus, the restriction (5.7) is
invariant under the all charging symmetry transformations.

In \cite {hk3} it was shown (in terms of $\X$ and $\A$) that the field
configuration (5.1) satisfies the
motion equations of the theory if it is restricted by Eq. (5.7). It
describes
the Israel--Wilson--Perj'es (IWP) class of solutions for the HS theory
reduced
to three dimensions. As the consequence of our consideration we conclude
that
{\it the IWP solutions form the CS invariant class}. This means
that
the complete CS subgroup $SO(2,d-1)\times SO(2,d-1+n)$ is the group of
invariance for the IWP solution. Moreover, if one interested in the arbitrary
asymptotics, one must perform the dressing transformations to
obtain the $U$--duality invariant class.

In the EM theory case the restriction (5.7) means that $\B=\exp {(i\beta)}$,
where $\beta$ is a real parameter. Using the usual procedure one can
compute
the NET and NHT generators; the result is:
\be
\Gamma_{{}_{NET}}=\partial_{\beta}, \quad
\Gamma_{{}_{NHT_1}}=\sin \beta \partial_{\beta}, \quad
\Gamma_{{}_{NHT_2}}=-\cos \beta \partial_{\beta}.
\ee
These generators also form the $su(1,1)$ algebra, its arising (instead of
the only expected $u(1)$ algebra) is related with the
non--trivial action of NHT
on the parameter $\beta$ {\it and} on the independent linearizing potential
$\Z_1$.
Thus, we have established that the IWP solutions form the CS invariant class
in the
EM and HS theories.

\setcounter{section}{0}
\setcounter{equation}{0}
\renewcommand{\theequation}{6.\arabic{equation}}
\section*{Generation Technique}
In a fact, starting from the arbitrary solution and performing {\it all}
the CS
(or U--duality) transformations one obtains the solution class which is
CS (or U--duality) invariant. Let us consider the {\it generation} of
the HS asimptoticaly--free solutions, which can be `dressed' in a usual
way. This process becomes simplest and manifestly CS--invariant in terms
of the linearizing potential $\Z$ (see Eq. (4.18)). The only question is
related to the choice of class of seed solutions.

It is natural to start with the $(d+3)$--dimensional Kaluza--Klein (KK)
theory. Being reduced to three dimensions, this theory is described by
the single matrix Ernst potential (see Eqs. (2.1)--(2.5))
\be
\X_{{}_{KK}}=
\left(
\ba{cc}
{\rm det}G&-u^T\cr
u&G
\ea
\right).
\ee
The $\X_{{}_{KK}}$ block elements can be combined into the
$SL(d+1,R)/SO(d+1)$ coset matrix \cite {cy}
\be
\M_{{}_{KK}}=
\left(
\ba{cc}
G^{-1}&\quad G^{-1}u\cr
&\cr
u^TG^{-1}&\quad {\rm det}G+u^TG^{-1}u
\ea
\right).
\ee
In its terms 
$\L_{{}_{KK}}=-\frac{1}{4}{\rm Tr}\,\left (\nabla \M_{{}_{KK}} \,\,
\nabla \M_{{}_{KK}}^{-1}\right )$, i.e. the KK theory obtains a
chiral form. This form allowed to obtain wide classes of the
three--dimensional solutions
defined by a set of harmonic functions (see \cite{kk}), etc.
Moreover, in two dimensions the inverse
scattering transform technique leads to construction of the
2N--soliton solution (see \cite {b} for the $d=1$ and $d=2$ cases). 

One can generate the HS theory solutions from the KK ones step--by--step.
Actually,
one can apply NST and NET to the seed solution $\Z_{1_{KK}}$
(constructed from
the known matrix $\X_{{}_{KK}}$). The result is the potential $\Z_{{1}_{B}}$, 
which contains not only the metric, but also the dilaton and Kalb--Ramond
fields. These are all the degrees of freedom of the bosonic string 
theory, i.e. {\it the normalized scaling and Ehlers transformations map
the Kaluza--Klein theory into the bosonic string one}. Next, an applying
of the NHT to the potential $\Z_{{1}_{B}}$ gives the whole sector of the HS
theory $\Z_{{1}_{HS}}$ and $\Z_{{2}_{HS}}$. This means that the {\it
normalized Harrison transformation maps the bosonic string theory into
the heterotic string one}. Finally, EMT mixes the $\Z_{{2}_{HS}}$ components.
Thus, we lead to the following {\it generation
technique}:
\be
\ba{ccccc}
\,&{}_{NET,\,\, NST}&\,&{}_{NHT,\,\, EMT}&\,\cr
Kaluza-Klein&\Longrightarrow&Bosonic\,\,\,String&\Longrightarrow&
Heterotic\,\,\,String\cr
\ea
\\
\nonumber
\ee

If the seed KK solution has the Coulomb term in its $1/r$--expansion, the
above described process simultaneously generates the charges:
NET and NST generate the dilaton and Kalb--Ramond charges, whereas NHT
and EMT generate the electric and magnetic ones. All these properties
explain the name `charging' for the symmetries preserving field asymptotics.
\setcounter{section}{0}
\setcounter{equation}{0}
\renewcommand{\theequation}{7.\arabic{equation}}
\section*{Concluding Remarks}
Thus, we have extracted all the charging symmetries from the U--duality
group of the effective three--dimensional theory of heterotic string.
We have established the linearizing potentials which undergo {\it linear
homogeneous transformations} when the charging symmetries act. We have
constructed one general invariant of these symmetries, quadratic on the
linearizing potentials.

It is shown that the charging symmetries preserve the {\it linear
homogeneous relation} between the linearizing potentials. It is
established that the
corresponding anzats defines the class of Israel--Wilson--Perj{'}es
solutions in the
special case when the above mentioned invariant vanishes. Furthermore,
the condition extracting IWP solution from the linear homogeneous anzats
occurs to be CS--invariant itself. Thus, it is stressed the close
relation between the charging symmetries, the linearizing potentials and
the IWP solutions.

It is well known that the IWP solutions describe supersymmetric field
configurations in any gravity theory which can be embedded into some
supergravity. For example, this takes place for the arbitrary EM
theory \cite {emsugra} and for the EM theory with the dilaton and axion
fields \cite {emdasugra}. In the forthcoming paper we hope to perform the
corresponding analysis for the IWP solution in the critical HS theory.

In the non--perturbative case the charges of field configuration become
quantized accordingly to the Dirac--Schwinger--Zwanziger rule \cite {dsz}.
The
IWP solution, being supersymmetric, does not obtain quantum corrections
to its charges, which become integer. The CS subgroup, as its group of
invariance, become the integer--valued group $SO(2,d-1;Z) \times
SO(2,d-1+n;Z)$.

The general charging symmetry invariant vanishes for the IWP solution
in the perturbative regime. The first non--trivial term in the CSI asymptotic
expansion near to the spatial infinity remains vanishing in the
non--perturbative regime as the BPS--bound for the supersymmetric solution.
It seems natural that CSI preserves its zero value in the whole
three--dimensional space for the supersymmetric IWP solutions in the
non--perturbative sector of HS theory.
\section*{Acknowledgments}
We thank our colleagues for encouraging us. A.H. was partially
supported by CONACYT.
\setcounter{section}{0}
\setcounter{equation}{0}
\renewcommand{\theequation}{A.\arabic{equation}}
\section*{Appendix A: LP for EM}
In this Appendix we derive the linearizing potentials of the
charging symmetry subgroup for the stationary Einstein--Maxwell theory.

The generators of the CS subgroup can be obtained from Eqs. (3.14)--(3.16);
the result is:
\be
\Gamma_{{}_{NHT_1}}\!=\!
F\left ( E\!-\!1\right )\partial_E\!+\!
\left ( F^2\!-\!E\!-\!1\right )\partial_F,
\quad
\Gamma_{{}_{NHT_2}}\!=\!-\!i\left [
F\left ( E\!-\!1\right )\partial_E\!+\!
\left ( F^2\!+\!E\!+\!1\right )\partial_F\right ].
\nonumber
\ee
\be
\Gamma_{{}_{NET}}=
i\left [ \left ( 1-E^2\right )\partial_E-
F\left ( E+1\right )\partial_F\right ],
\quad
\Gamma_{{}_{EMT}}=iF\partial_F,
\ee
Here $\Gamma_{{}_{NHT_1}}$ $(\Gamma_{{}_{NHT_2}})$ corresponds to the real
(imaginary) part of the Harrison's parameter $\lambda_{\H}$. It is useful
to introduce the new set of generators
\be
X_1=-\frac {1}{2\sqrt 2}\Gamma_{{}_{NHT_1}}, \quad
X_2=\frac {1}{2\sqrt 2}\Gamma_{{}_{NHT_2}}, 
\ee
\be
X_3=\frac {1}{4}\left (2\Gamma_{{}_{EMT}}-\Gamma_{{}_{NET}}\right ), \quad
X_4=\frac {1}{4}\left (2\Gamma_{{}_{EMT}}+\Gamma_{{}_{NET}}\right ); 
\ee
then 
\be
\left [ X_1,X_2\right ]=-X_3,\quad \left [ X_2,X_3\right ]=X_1,\quad
\left [ X_3,X_1\right ]=X_2,
\ee
and $\left [ X_1,X_4\right ]=\left [ X_2,X_4\right ]=
\left [ X_3,X_4\right ]=0$. Thus, three generators $X_1$, $X_2$ and $X_3$
form the $su(1,1)$ subalgebra, whereas the generator $X_4$ defines the
commuting subalgebra $u(1)$.

All the CS transformations can be realized $linearly$ in the following way.
Let $\Z=(\Z_1, \Z_2)$ be the complex $2\times 1$ row whose finite
transformation is:
\be
\Z\rightarrow\G^{{}^{left}}\Z\G^{{}^{right}}.
\ee
We define $\G^{{}^{left}}$ as the $U(1)$ group factor, and
$\G^{{}^{right}}$ as the $SU(1,1)$ group matrix:
\be
\G^{{}^{left}}\bar\G^{{}^{left}}=1, \quad 
\G^{{}^{right}}\sigma_3\left (\G^{{}^{right}}\right )^+=\sigma_3.
\ee
Then in the infinitesimal case $\G^{{}^{left}}=1+i\gamma^4$ and
$\G^{{}^{right}}=1+\sum \gamma^i\tau_i$, where the $SU(1,1)$ matrix
generators are:
\be
\tau_1\!=\!\frac {1}{2}\sigma_1\!=\!\frac {1}{2}
\left(
\ba{cc}
0&1\cr
1&0
\ea
\right), \quad
\tau_2\!=\!\frac {i}{2}\sigma_2\!=\!\frac {i}{2}
\left(
\ba{cc}
0&-1\cr
1&0
\ea
\right), \quad
\tau_3\!=\!-\frac {i}{2}\sigma_3\!=\!-\frac {i}{2}
\left(
\ba{cc}
1&0\cr
0&-1
\ea
\right). 
\ee
The corresponding functional form of the generators is:
\be
&&Y_1=\frac {1}{2}\left (Z_2\partial_1+Z_1\partial_2\right ), \quad
Y_2=\frac {i}{2}\left (Z_2\partial_1-Z_1\partial_2\right ),
\nonumber\\
&&Y_3=\frac {i}{2}\left (Z_2\partial_2-Z_1\partial_1\right ), \quad
Y_4=\frac {i}{2}\left (Z_1\partial_1+Z_2\partial_2\right ),
\ee
where $Y_4$ is the generator for $\G^{{}^{left}}$.
One can prove that they satisfy the same commutation relations as $X$'s.

Now we identify the $X$--generators and the $Y$--generators with the
equal subscripts. Supposing that the functional relations
$E=E(\Z_1, \Z_2)$ and $F=F(\Z_1, \Z_2)$ exist, we obtain the differential
equations of the first order which define them. Our plan is
following: we will consider the differential equations arising from the two
identifications $Y_1\pm iY_2=X_1\pm iY_2$. Then $Y_3=X_3$ will take place
in view of the $su(1,1)$--algebra commutation relations, and the last
equality $Y_4=X_4$ will be satisfied because we have normalized these
generators
in a `right' way (in the other case they will be proportional).

Let us briefly consider the solution process. The `$\pm$'-identifications
have the form:
\be
&&\sqrt 2\Z_1\partial_2=\left (E+1\right )\partial_F,
\\
&&\sqrt 2\Z_2\partial_1=-F\left (E-1\right )\partial_E-F^2\partial_F.
\ee
Eq. (A.9) is equal to
\be
E,_2=0, \quad \sqrt 2\Z_1F,_2=E+1.
\ee
From this we conclude that $E=E(\Z_1)$ and $F=\frac {1}{\sqrt 2}
\frac {\Z_2}{\Z_1}\left ( E+1\right ) + \psi (\Z_1)$, where $\psi$ is the
arbitrary function. Next, Eq. (A.10) is consistent if $\psi =0$;
the following integration leads to
\be
E=\frac {2}{C\Z_1-1}+1, \quad F=\sqrt 2\,\frac {C\Z_2}{C\Z_1-1},
\ee
where $C$ is the arbitrary constant. It is obvious that the linearizing
potentials are defined at least up to multiplier. Then, using the
replacement $C\Z\rightarrow\Z$, and inversing the formulae (A.12) we obtain
that
\be
\Z_1=\frac {2}{E-1}+1, \quad \Z_2=\sqrt 2\,\frac {F}{E-1}.
\ee
These potentials had been introduced at the first time tby Mazur in \cite {m}
without any relation to the CS subgroup linearization problem.
\setcounter{section}{0}
\setcounter{equation}{0}
\renewcommand{\theequation}{B.\arabic{equation}}
\section*{Appendix B: CS Algebra for HS}
In this Appendix we compute the commutation relations for the charging
symmetry algebra of the heterotic string theory with arbitrary $d$ and $n$.

First, the generators of the subgroup $\G^{{}^{left}}$ form the $so(2,d-1)$
algebra in
the usual way. Let us consider the generators $\Gamma^{{}^{right}}$ of
the $so(2,d-1+n)$ algebra (see Eq. (4.17)).
They are constructed in the infinitesimal form; this means
that $\lambda_{\E}=-2\xi e$, $\lambda_{\H}=\sqrt 2\xi h$ and
$\tau =\xi t$, where
$\xi$ is the infinithesimal parameter. The matrices $e$, $h$ and $t$
are finite ($e^T=-e$, $t^T=-t$); they define the finite form of the NET, NHT
and EMT generators:
\be
\Gamma_{{}_{NET}}\left ( e\right )=
\left(
\ba{cc}
\Sigma e&0\cr
0&0
\ea
\right),
\quad
\Gamma_{{}_{NHT}}\left ( h\right )=
\left(
\ba{cc}
0&\Sigma h\cr
-h^T&0
\ea
\right),
\quad
\Gamma_{{}_{EMT}}\left ( t\right )=
\left(
\ba{cc}
0&0\cr
0&t
\ea
\right).
\ee
The computation of the `mixing' commutators gives:
\be
&&\left [ \Gamma_{{}_{EMT}}\left ( t\right ),
\Gamma_{{}_{NET}}\left ( e\right )\right ]
=0,
\nonumber\\
&&\left [ \Gamma_{{}_{NET}}\left ( e\right ),
\Gamma_{{}_{NHT}}\left ( h\right )\right ]
=\Gamma_{{}_{NHT}}\left ( e\Sigma h\right ),
\nonumber\\
&&\left [ \Gamma_{{}_{NHT}}\left ( h\right ),
\Gamma_{{}_{EMT}}\left ( t\right )\right ]
=\Gamma_{{}_{NHT}}\left ( ht\right ).
\ee
Next, for the `inner' commutation relations one obtains:
\be
&&\left [ \Gamma_{{}_{EMT}}\left ( t_1\right ),
\Gamma_{{}_{EMT}}\left ( t_2\right )\right ]
=\Gamma_{{}_{EMT}}\left ( t_{[1}t_{2]}\right ),
\nonumber \\
&&\left [ \Gamma_{{}_{NET}}\left ( e_1\right ),
\Gamma_{{}_{NET}}\left ( e_2\right )\right ]
=\Gamma_{{}_{NET}}\left ( e_{[1}\Sigma e_{2]}\right )
\nonumber\\
&&\left [ \Gamma_{{}_{NHT}}\left ( h_1\right ),
\Gamma_{{}_{NHT}}\left ( h_2\right )\right ]
=-\Gamma_{{}_{NET}}\left ( h_{[1}h_{2]}^T\right )-
\Gamma_{{}_{EMT}}\left ( h_{[1}^T\Sigma t_{2]}\right ).
\ee
Thus, the NET and EMT subalgebras commute, whereas the commutator of NHT
with NET and EMT is NHT again. However, NHT does not form a subalgebra
with NET or EMT, because from the `inner' commutation relations it follows
that the
minimal algebra including NHT is equal to the full `right' sector of the
charging symmetry algebra, i.e. to $so(2,d-1+n)$.

\end{document}